\documentclass{amsart}
\usepackage{amssymb}
\usepackage{latexsym}
\newcommand{\Z}{{\mathbb Z}}
\newcommand{\R}{{\mathbb R}}
\newcommand{\C}{{\mathbb C}}
\newcommand{\N}{{\mathbb N}}

\newcommand{\beq}{\begin{eqnarray}}
\newcommand{\eeq}{\end{eqnarray}}

\newtheorem{theorem}{Theorem}

\newtheorem{lemma}{Lemma}[section]
\newtheorem{prop}[lemma]{Proposition}
\newtheorem{coro}[lemma]{Corollary}

\newcommand{\tr}{{\mathrm{tr}}}  
\def\pp{\phi}
\def\e{\varepsilon}

\begin{document}
\title[Quantum Dynamics in One Dimension]{Power-Law Bounds on Transfer Matrices and Quantum Dynamics in One Dimension}
\author[D.\ Damanik, S.\ Tcheremchantsev]{David Damanik$\, ^1$ and Serguei Tcheremchantsev$\, ^2$}
\thanks{D.\ D.\ was supported in part by NSF Grant No.~DMS--0010101}
\maketitle
\vspace{0.3cm}
\noindent
$^1$ Department of Mathematics 253--37, California Institute of Technology, Pasadena, CA 91125, USA\\[2mm]
$^2$ UMR 6628 -- MAPMO, Universit\'{e} d'Orleans, B.P.~6759, F-45067 Orleans C\'{e}dex, France\\[3mm]
E-mail: \mbox{damanik@its.caltech.edu, serguei.tcherem@labomath.univ-orleans.fr}\\[3mm]

\begin{abstract}
We present an approach to quantum dynamical lower bounds for discrete one-dimensional Schr\"odinger operators which is based on power-law bounds on transfer matrices. It suffices to have such bounds for a nonempty set of energies. We apply this result to various models, including the Fibonacci Hamiltonian.
\end{abstract}

%
%
%
%

\section{Introduction}

Consider a self-adjoint operator $H$ on a separable Hilbert space $\mathcal{H} = \ell^2 (\Z^d)$ or $\ell^2 (\N)$. The dynamical evolution of an initial state $\psi \in \mathcal{H}, \ \Vert \psi \Vert =1$, is given by $\psi(t)=\exp (-itH) \psi$. We shall denote
$\psi(t,n)=\langle \psi(t), \delta_n \rangle$, where $\mathcal{B}=\{ \delta_n \}$ is the canonical basis of $\mathcal{H}$ labelled by $n \in \Z^d$ or $n \in \N$. One usually takes $\psi$ so that $\psi(0,n)=\psi(n)$ be well localized (fast decaying at infinity). For example, one can take $\psi=\delta_1$ in the one-dimensional case. While being localized at $t=0$, the wave packet in general spreads with time over the basis $\mathcal{B}$. It is convenient to consider the time-averaged quantities

\begin{equation}\label{ant}
a(n,T)=\frac{1}{T} \int_0^T |\psi(t,n)|^2 \, dt,  \ \
\sum_n a(n,T)=\Vert \psi \Vert ^2=1 \ {\rm for} \ {\rm all} \ T>0. 
\end{equation}
There exist basically two possibilities to characterize the spreading of the wave packet. First, one can consider the upper and lower rates associated to the fastest (or the slowest) part of the wave packet. Let
$$
S(\gamma,T)=\sum_{n: |n|\ge T^\gamma-2} a(n,T), \ \gamma \ge 0. 
$$
For the fastest part, one defines

\begin{equation}\label{in1}
\gamma^- = \sup \left\{ \gamma \ge 0 \ | \ \liminf_{T \to +\infty} \frac{{\rm log} \, S(\gamma,T) }{ {\rm log} \, T } = 0 \right\},
\end{equation}

\begin{equation}\label{in2}
\gamma^+=\sup \left\{ \gamma \ge 0 \ | \ \limsup_{T \to +\infty} \frac{ {\rm log} \, S(\gamma,T) }{ {\rm log} \, T } = 0 \right\}
\end{equation}
(with convention ${\rm log} \, 0 = -\infty$).
Assume that $\gamma^\pm>0$. Then it follows from this definition
that for any $0<\nu<\gamma^-, \eta>0$,

\begin{equation}\label{in01}
\sum_{n: |n|\ge T^{\gamma^- -\nu}} a(n,T) \ge C T^{-\eta}
\end{equation}
for all $T\ge 1$ with a uniform (in $T$) constant $C$. For $\gamma^+$, a
similar bound holds on some sequence of times $T_k \to +\infty$.
One could take a slightly different definition of $\gamma^\pm$ so that
one has for any $0<\nu<\gamma^-$, 
\begin{equation}\label{in010}
\sum_{n: |n| \ge T^{\gamma^- -\nu}} a(n,T) \ge C(\nu)>0
\end{equation}
(and similarly for $\gamma^+$), but the meaning of numbers $\gamma^{\pm}$ will be essentially the same.

The spreading rates for the slowest part of the wave packet can be defined with summation over $\{ n: |n|\le T^{\gamma} \}$ and taking $\inf \{ \gamma \}$ in \eqref{in1}, \eqref{in2} (see also \cite{kkl} for a slightly different definition). In the present paper we shall be interested, however, only in the fastest part of the wave packet. 

To determine the numbers $\gamma^\pm$ for concrete quantum systems, one should be able to relate it to the spectral properties of $H$. What is in fact known are some general lower bounds for $\gamma^\pm$. Although not stated in this form, it follows from the proofs of \cite{bcm,l} that

\begin{equation}\label{in3}
\gamma^- \ge \frac{1}{d} {\rm dim}_H(\mu_\psi), 
\end{equation}
and from the proof of \cite{gs} that

\begin{equation}\label{in4}
\gamma^+ \ge \frac{1}{d} {\rm dim}_P(\mu_\psi) 
\end{equation}
(with $d \ge 1$ in the case of $l^2(\Z^d)$ and $d=1$ in the case
of $l^2(\N)$). 
Here $\mu_\psi$ is the spectral measure associated to the state $\psi$ and the operator $H$, and ${\rm dim}_H(\mu), \ {\rm dim}_P(\mu)$ denote the (upper) Hausdorff and packing dimensions of the measure $\mu$, respectively. For different definitions and relations between them, see, for example, \cite{bgt2,f,t}. In particular, the Hausdorff dimension is determined by the most continuous part of the measure $\mu$:
$$
{\rm dim}_H(\mu)= \mbox{$\mu$-ess-sup} \, \lambda^-(E), 
$$
where $\lambda^-(E)$ is the lower local exponent of $\mu$,
$$
\lambda^-(E)=\liminf_{\e \to 0} \frac{ {\rm log} \, \mu([E-\e,E+\e]) }{ {\rm log} \, \e}, \ \ E \in {\rm supp} \, \mu. 
$$
For the packing dimension, we have
$$
{\rm dim}_P(\mu)= \mbox{$\mu$-ess-sup} \, \lambda^+(E),
$$
where $\lambda^+(E)$ is the upper local exponent of the measure,
$$
\lambda^+(E)=\limsup_{\e \to 0} \frac{ {\rm log} \, \mu([E-\e,E+\e]) }{ {\rm log} \, \e}, \ \ E \in {\rm supp} \, \mu. 
$$
One can observe that for Hausdorff dimension, the result is slightly stronger than (\ref{in3}), namely, (\ref{in010}) holds with $\gamma^-={\rm dim}_H(\mu_\psi)$. 

If one knows the continuity properties of the spectral measure (encoded in $\lambda^-(E)$) and one has also some information about the decay of the generalized eigenfunctions $u_\psi(n,E)$, associated to the state $\psi$, then one can improve the lower bound \eqref{in3} for $\gamma^-$ \cite{kkl,kl}.  

More involved quantities which describe the fastest portion of the wave packet are the time-averaged moments of the position operator:

\begin{equation}\label{mpo}
\langle |X|_\psi^p \rangle (T)=\frac{1}{T} \int_0^T \sum_n |n|^p |\psi(t,n)|^2 =\sum_n |n|^p a(n,T), \ p>0 
\end{equation}
(where $|n|$ is the norm in $\Z^d$ in the case of $l^2(\Z^d)$). One can define the associated upper and lower growth exponents,
$$
\beta^-(p)=\liminf_{T \to +\infty} \frac{{\rm log} \, \langle |X|_\psi^p \rangle (T) }{ {\rm log} \, T},
$$
and in a similar manner for $\beta^+(p)$. The numbers $\beta^\pm(p)$ depend, in general, on the state $\psi$, but we will leave the dependence implicit. It is clear that \eqref{in01} implies
$$
\langle |X|_\psi^p \rangle (T) \ge C T^{p(\gamma^--\nu)-\eta}
$$
for any $\nu>0, \eta>0$ uniformly in $T\ge 1$ and thus $\beta^-(p) \ge p \gamma^-$. Together with (\ref{in3}) this yields

\begin{equation}\label{in4b}
\beta^- (p) \ge \frac{p}{d} {\rm dim}_H(\mu_\psi),  
\end{equation}
the bound most often used to bound from below the moments of the
position operator. 
Similarly, one always has
$\beta^+(p) \ge p \gamma^+\ge \frac{p}{d} {\rm dim}_P (\mu_\psi)$.

It is important to observe that strict inequalities $\beta^-(p)>p \gamma^-, \ \beta^+(p)>p\gamma^+$ may occur. This is possible if the wave packet has polynomially decaying tails at infinity. Assume, for example, that for some $\gamma >\gamma^-$ and some $\tau>0$,

\begin{equation}\label{in5}
\sum_{n: |n| \ge T^\gamma} a(n,T) \ge C T^{-\tau}
\end{equation}
uniformly in $T\ge 1$. Then $\beta^-(p)\ge p \gamma -\tau$. This bound is better than $\beta^-(p) \ge p \gamma^-$ for $p$ large enough.

General lower bounds for the time-averaged moments which take into account (in a somewhat hidden form) polynomial tails, are obtained in \cite{bgt1} and \cite{t}. The proofs are based on the spectral theorem and develop the ideas of Guarneri \cite{g3}. The obtained lower bounds are expressed in terms of spectral measure $\mu_\psi$ \cite{bgt1,t} or in terms of both spectral measure and generalized eigenfunctions $u_\psi (n,E)$ \cite{t}. To apply them to concrete quantum systems, one should have a good knowledge of $\mu_\psi ([E-\e,E+\e]), \ E \in {\rm supp} \,  \mu_\psi$ (and also of the functions $S_N(E)=\sum_{n: |n| \le N} |u_\psi (n,E)|^2$ in the case of the mixed lower bounds from \cite{t}). Such a kind of information is difficult to obtain in the multidimensional case. However, in one dimension for operators $H$ of the form 

\begin{equation}\label{oper}
H \psi(n)=\psi(n-1)+\psi(n+1) + V(n)\psi(n),
\end{equation}
there exist rather effective methods based on the study of solutions to the formal eigenfunction equation $Hu=Eu, \ E \in \R$. In particular, the growth properties of the transfer matrix $T(n,E)$ associated with this equation are closely related with the spectral properties of operator $H$. For simplicity, from now on we shall always take $\psi = \delta_1$ and we write simply $\mu$ instead of $\mu_{\delta_1}$.

Our purpose is not to give a detailed list of results obtained in this field. We will mainly be interested in models with power-law growth of the norm of the transfer matrix. Roughly speaking, the slower the growth of $\|T(n,E)\|$, for $E \in A$ and as $|n| \to \infty$, the more continuous is the restriction of $\mu$ on $A$. What will be of particular interest for us in the present paper, are the two following results. First, if $\|T(n,E)\| \le C(E) |n|^\eta$ for $\mu$-a.e.\ $E \in A, \ \mu(A)>0$ with $\eta \in [0,1/2)$ and finite $C(E)$, then \cite{jl1,jl2}
$$
\mu([E-\e,E+\e]) \le D(E) \e^{1-2\eta}, \  E \in A, 
$$
so that the measure is $(1-2\eta)$-continuous on $A$. In particular,
\begin{equation}\label{in51}
{\rm dim}_H(\mu) \ge 1-2\eta, \ \ \beta^-(p) \ge p(1-2 \eta).  
\end{equation}

More generally, assume that every solution of the equation $Hu=Eu, \ E \in A$ obeys

\begin{equation}\label{upperandlower}
C_1(E) L^{2q_1} \le \sum_{n: |n| \le L} |u(n,E)|^2 \le C_2(E) L^{2q_2}
\end{equation}
with positive finite $C_1(E), C_2(E)$ and $0<q_1 \le q_2 <+\infty$. Then the measure is $2q_1/(q_1+q_2)$-continuous on $A$ \cite{dkl,jl2}. In particular, ${\rm dim}_H(\mu) \ge 2q_1/(q_1+q_2)$. One can observe that the condition $\|T(n,E)\| \le C(E) |n|^\alpha$ with some $\alpha \ge 0$ implies $q_2 \le \alpha+1/2$. 

The polynomial upper bound on the norm of the transfer matrix also allows one to bound from \textit{below} $\mu([E-\e,E+\e])$. Such a bound can be used, following general results of \cite{bgt1,t}, to obtain non-trivial dynamical bounds for the moments ($\beta^-(p)>0$) for some systems with pure point spectrum \cite{gkt}. Since in this case ${\rm dim}_H(\mu)=0$, these bounds cannot be obtained by usual methods based on the continuity properties of the measure. (Another example of this kind, studied by Jitomirskaya et al.\ in \cite{jss}, will be discussed later on).   

As was said above, the general results of \cite{bgt1,t} were obtained using the spectral theorem for the operator $H$ and thus by representing all the quantities of interest as some integrals over the spectral measure. In the present paper we propose another method based on integrals over Lebesgue measure rather than over $\mu_\psi$. One can bound from below
rather directly the sums $\sum_{n: |n| \ge L} a(n,T)$
(and thus the time-averaged moments $\langle |X|^p \rangle (T)$) using
the Parseval equality \cite{RS4}. Namely, for any $\psi \in \mathcal{H}$,

\begin{equation}\label{in6}
\frac{1}{T} \int_0^{+\infty} e^{-2t/T} |\psi(t,n)|^2 \, dt =
\frac{1}{2 \pi T} \int_{\R} |(R(E+i/T)\psi)(n)|^2 \, dE.
\end{equation}
Here, $n \in \Z^d$ or $n \in \N$, and $R(z)=(H-zI)^{-1}$ is the resolvent of $H$. The integral on the l.h.s.\ in \eqref{in6} is very close to the Cesaro time-averaged quantity $a(n,T)$. Therefore we modify the definition of $a(n,T)$ in \eqref{ant} and of $\langle |X|^p \rangle (T)$ in \eqref{mpo} accordingly (this is done in this way by many authors),

\begin{equation}\label{antmpomod}
a(n,T)=\frac{1}{T} \int_0^{+\infty} e^{-2t/T} |\psi(t,n)|^2 \, dt , \; \langle |X|_\psi^p \rangle (T) = \sum_n |n|^p a(n,T),
\end{equation}
and work with these definitions in what follows. One can easily see that this does not change the numbers $\beta^\pm(p)$, provided that the moments $|X|^p_\psi (t)$ do not grow faster than polynomially (which is true in most applications and, in particular, in all applications considered in the present paper). 

To obtain a non-trivial dynamical lower bound, it is sufficient to bound from below $\sum_{n: |n| \ge T^\gamma} |R(E+i/T)\psi (n)|^2$ for some $\gamma>0$ for $E$ from some set of positive Lebesgue measure. In our one-dimensional case with $\psi=\delta_1$, the resolvent $R(E+i/T)\delta_1(n)$ is closely related to the corresponding transfer matrix $T(n,1;E+i/T)$, which is in turn close to $T(n,1;E)$ if $T$ is large and $n$ is not too large. Roughly speaking, if $\|T(n,1;E)\|$ is bounded from above polynomially in $n$ for some $E$, then $R(E+i/T)\delta_1(n)$ decays no faster than polynomially for $n$ not too large (depending on $T$). The following result will be proved in Section~2:

\begin{theorem}\label{main}
Let the operator $H$ in $\ell^2(\Z)$ or $\ell^2(\N)$ be given by \eqref{oper} {\rm (}with Dirichlet boundary conditions at $0$ in the half-line case{\rm )}. Suppose that for some $K>0$, $C>0$, $\alpha >0$, the following condition holds:

For any $N>0$ large enough, there exists a nonempty Borel set $A(N)\subset \R$ such that 
$A(N) \subset [-K,K]$ and 

\begin{equation}\label{maincond}
\|T(n,m;E)\| \le C N^{\alpha} \ \  \forall E \in A(N),\ \forall \ n,m: |n|\le N, |m| \le N
\end{equation}
{\rm (}resp., with $1 \le n \le N, \ 1 \le m \le N$ in the case of $\ell^2(\N)${\rm )}. 

Let $N(T)=T^{1/(1+\alpha)}$ and let $B(T)$ be the $1/T$-neighborhood of the set $A(N(T))$:
$$
B(T)=\{ E \in \R : \exists E' \in A(N(T)), |E-E'| \le 1/T \}.
$$

Then for all $T > 1$ large enough, the following bound holds:

\begin{equation}\label{f8}
\sum_{|n| \ge N(T)/2} a(n,T) \ge \frac{\hat{C}}{T} |B(T)| N^{1-2\alpha} (T), 
\end{equation}
where $\hat{C}$ is some uniform positive constant and
$|B|$ denotes the Lebesgue measure of the set $B$.

In particular, for any $p > 0$, one has the following bound for the time-averaged moments of 
the position operator:

\begin{equation}\label{f9}
\langle |X|_\psi^p \rangle (T) \ge \frac{\hat{C}}{T}|B(T)| N^{p+1-2 \alpha} (T),
\end{equation}
where $\langle |X|_\psi^p \rangle (T)$ is defined as in \eqref{antmpomod}.
\end{theorem}

In the statement of Theorem~\ref{main} it is important that the constant $C$ in \eqref{maincond} be independent of $N$ and of $E \in A(N)$. If $A(N)$ consists of a single point $E_0$ (independently of $N$), the statement of the theorem is still non-trivial. Namely, 

\begin{coro}[One-Energy Theorem]\label{oneenergy}
If

\begin{equation}\label{oneecond}
\|T(n,m; E_0)\| \le C(E_0)(|n|+|m|)^{\alpha}
\end{equation}
for some $E_0$, uniformly in $n,m$, then
$$
\beta^-(p) \ge \frac{p-1-4 \alpha}{1+\alpha}.
$$
\end{coro}
If one only knows that $\|T(n,1; E_0)\| \le C(E_0) |n|^\eta$, then one can take $\alpha = 2 \eta$, so that

\begin{equation}\label{in7}
\beta^-(p) \ge \frac{p-1-8 \eta}{1+2 \eta}. 
\end{equation}
It is interesting to compare this bound with \eqref{in51}, stating for $0 \le \eta <1/2$, that $\beta^-(p) \ge p(1-2 \eta)$. One observes that \eqref{in7} yields a better result for $p$ large enough (and one needs the polynomial upper bound for the transfer matrix only for a single point
and not on some set of positive measure). Moreover, in the case $\eta \ge 1/2$, the bound \eqref{in7} is always non-trivial for $p$ large enough, while \eqref{in51} gives nothing. In this case the spectrum of $H$ may be pure point with polynomially decaying eigenfunctions. 

Another interesting consequence of Theorem~\ref{main} is the following:

\begin{coro}\label{coro2}
If there exist $C>0, \ E_0 \in \R, \ \theta \ge 1$ such that
$$
\|T(n,1; E)\| \le C 
$$
for all $n,E$ such that $|n| \cdot |E-E_0|^{\theta} \le 1$, then
$$
\beta^-(p) \ge p-1/\theta. 
$$
\end{coro}
Taking $\theta=2$ in Corollary~\ref{coro2}, we obtain the main dynamical result of \cite{jss}.

Finally, we note that all the results mentioned above are stable with respect to finitely supported perturbations. This is of interest since all the previous dynamical lower bounds were based on dimensionality properties of spectral measures which are very sensitive to such perturbations.

\begin{coro}\label{coro3}
Suppose the potential $V_0$ is such that the operator $H_0$ with potential $V_0$ satisfies the hypothesis of Theorem~\ref{main}. Let $W : \Z \rightarrow \R$ {\rm (}resp., $W: \N \rightarrow \R${\rm )} be a finitely supported perturbation and $H = H_0 + W$. Then the operator $H$ satisfies \eqref{f8} and \eqref{f9}. The same kind of stability holds for Corollaries~\ref{oneenergy} and \ref{coro2}.
\end{coro}

We shall apply the results above to various models from one-dimensional quasicrystal theory (cf.~\cite{d2}): the Fibonacci Hamiltonian, the period doubling Hamiltonian, and the Thue-Morse Hamiltonian. More applications will be discussed in a forthcoming publication \cite{dst}.

The Fibonacci potential $V : \Z \rightarrow \R$ is given by

\begin{equation}\label{fibpot}
V(n) = \lambda \chi_{[1- \omega, 1)} (n \omega \mod 1).
\end{equation}
Here, $\lambda > 0$ and $\omega$ is the inverse of the golden mean, that is, $\omega = (\sqrt{5} - 1)/2$. For a given $\lambda$, let 

\begin{equation}\label{clambda}
C_\lambda = 2 + \sqrt{8 + \lambda^2}.
\end{equation}

We shall prove

\begin{theorem}\label{fibtheo}
{\rm (a)} For every $\lambda, p$, we have
$$
\beta^- (p)  \ge \frac{p - 1 - 4 \alpha}{ 1 + \alpha}, 
$$
with 

\begin{equation}\label{alphadef}
\alpha = \alpha(\lambda) = \frac{2 \log (C_\lambda (2 C_\lambda + 1)^2)}{\log \omega^{-1}}.
\end{equation}
{\rm (b)} For every $p$ and every $\lambda > 4$, we have
$$
\beta^- (p) \ge \frac{p - \gamma - 3 \alpha}{ 1 + \alpha}, 
$$
where $\alpha$ is as in \eqref{alphadef} and 

\begin{equation}\label{gammalambda}
\gamma = \gamma (\lambda) =
\frac{\log (2 \lambda + 22)}{\log \omega^{-1}} - 1 <1+\alpha.
\end{equation}
\end{theorem}

Let us compare the result of Theorem~\ref{fibtheo} with previously known lower bounds for the moments. In fact, all these bounds are based on \eqref{in4b}. The lower bounds for the Hausdorff dimension for the Fibonacci Hamiltonian \cite{d1,jl2,kkl} in turn have all been proved using \eqref{upperandlower}. For example, in \cite{kkl} it was proved that

\begin{equation}\label{fibzerophase}
{\rm dim}_H (\mu) \ge 2 \kappa/(\kappa + \alpha+1/2),
\end{equation}
where
$$
\kappa = \frac{\log (\sqrt{17}/4)}{5 \log( \omega^{-1} )} \approx 0.0126
$$
and $\alpha$ is given by \eqref{alphadef}. 
For large $p$, the dominant expression in the lower bound for $\beta^-(p)$
in Theorem~\ref{fibtheo} is $p/(1+\alpha)$ which is clearly better than
$2\kappa p/(\kappa+\alpha+1/2)$. We therefore see that our result improves the previously known lower bound for large $p$.

We also note that Corollary~\ref{coro3} is particularly interesting in this context. The Fibonacci Hamiltonian has spectrum $\sigma$ of zero Lebesgue measure \cite{bist,s2} and hence, by a suitable application of the Simon-Wolff argument \cite{sw}, a generic rank two perturbation
(at two consecutive sites) will produce a spectral measure which is entirely supported away from $\sigma$. More precisely, the spectral measure of the perturbed operator will be supported on a countable set of eigenvalues with $\sigma$ being the set of accumulation points of these
eigenvalues. In particular, a generic rank two perturbation may turn the spectral measure of the Fibonacci Hamiltonian, which is known to be $\alpha$-continuous for some $\alpha>0$, into a pure point measure. In this situation, previous bounds do not give anything for the perturbed operator, while our bound for $\beta^-(p)$ is stable with respect to such a perturbation.

Our next application is to the period doubling Hamiltonian which was considered, for example, in \cite{bbg,bg,d0,d3,hks}. Let $\Omega_{{\rm pd}}$ be the (two-sided) subshift generated by the substitution $0 \mapsto 01$, $1 \mapsto 00$ and define, for some $\omega \in \Omega_{{\rm pd}}$ and a given coupling constant $\lambda$, the potential $V_{\lambda,\omega}$ by $V_{\lambda,\omega} (n) = \lambda \omega_n$. Is is known that for every $\lambda \not=0$ and every $\omega \in \Omega_{{\rm pd}}$, the operator with potential $V_{\lambda,\omega}$ has purely singular continuous spectrum (see \cite{d3} for this result and \cite{bbg,bg,d0,hks} for earlier partial results). We shall show the following:

\begin{theorem}\label{pdtheo}
For every $\lambda$ and every $\omega \in \Omega_{{\rm pd}}$, we have
$$
\beta^- (p) \ge \frac{p - 5}{2}.
$$
\end{theorem} 

This result is of interest for various reasons. First of all, this is the first dynamical result for this model that goes beyond the RAGE theorem (i.e., the dynamical result that follows from purely continuous spectrum). Indeed, apart from singular continuity, nothing on either dimensionality or dynamics was known rigorously. The difference to the Fibonacci case, where, as discussed above, earlier results in these directions were known, stems from the absence of an invariant for the trace map. More concretely, it could not be shown that the trace map orbits remain bounded for (sufficiently many) energies from the spectrum. With a result like Corollary~\ref{oneenergy} at our disposal, very weak results on the period doubling trace map suffice already to allow us to establish the result above. Secondly, observe that the dynamical bound we obtain in Theorem~\ref{pdtheo} is independent of the coupling constant. This is in sharp contrast to the Fibonacci case where our bound (and previous ones) becomes worse as the coupling constant is increased.

Finally, we consider the Thue-Morse Hamiltonian which was studied, for example, in \cite{ap,bel,bg,dp,hks}. Let $\Omega_{{\rm tm}}$ be the (two-sided) subshift generated by the substitution $0 \mapsto 01$, $1 \mapsto 10$ and define as above, for some $\omega \in \Omega_{{\rm tm}}$ and a given coupling constant $\lambda$, the potential $V_{\lambda,\omega}$ by $V_{\lambda,\omega} (n) = \lambda \omega_n$. We shall show the following:

\begin{theorem}\label{tmtheo}
For every $\lambda$ and every $\omega \in \Omega_{{\rm tm}}$, we have
$$
\beta^- (p) \ge p - 1.
$$
\end{theorem} 

As in the previous theorem, the bound is $\lambda$-independent and the first dynamical result for this model. Furthermore, in this case the spectral type has not even been identified in all cases. It is known that for every $\lambda \not= 0$ and every $\omega \in \Omega_{{\rm tm}}$,
the absolutely continuous spectrum is empty (this follows from results of Kotani \cite{k} and Last and Simon \cite{ls}), but absence of eigenvalues is only known generically, that is, for a
dense $G_\delta$ set of $\omega \in \Omega_{{\rm tm}}$ (this was shown by Delyon and Peyri\`{e}re \cite{dp} and Hof et al.\ \cite{hks}, using different methods). This illustrates the theme of this paper in a nice way: While we cannot say anything about the
dimensionality of spectral measures, we can nevertheless prove very strong dynamical bounds.

The organization of the article is as follows: Theorem~\ref{main} and Corollaries~\ref{oneenergy}--\ref{coro3} will be proved in Section~2. In Sections~3--5 we shall then apply these criteria to the Fibonacci model, the period doubling model, and the Thue-Morse model, respectively.

%
%
%
%

\section{A New Criterion for Quantum Dynamical Lower Bounds}
 
Let $H=-\Delta+V$ be a self-adjoint operator on $\ell^2(\Z)$ or on $\ell^2(\N)$ (with Dirichlet boundary condition in the latter case). Given some real function $V(n)$, 
it is formally defined by
$$
H \psi(n)=\psi(n-1)+\psi(n+1) +V(n)\psi(n), \ \ \ n \in \Z
$$
and by
$$
H \psi(n)=\psi(n-1)+\psi(n+1) +V(n)\psi(n), \ \ \ n>1, 
$$
$$
H \psi(1)=\psi(2) +V(1)\psi(1) 
$$
respectively. Let $z$ be some complex or real number. We define the transfer matrix associated to the operator
$H$ as follows:

\begin{equation}\label{transfermat}
T(n,m;z) = \left\{ \begin{array}{cl} A(n,z)A(n-1,z) \cdot ... \cdot A(m+1,z) & \mbox{ if } n > m, \\ I & \mbox{ if } n=m, \\ T^{-1}(m,n;z) & \mbox{ if } n < m, \end{array} \right. 
\end{equation}
where 
$$
A(n,z)=\left( \begin{array}{cr} z-V(n) &  -1 \\ 1 & 0 \end{array} \right).
$$
For $z \in \C$ with ${\rm Im} \, z \ne 0$, define
$$
\pp=R(z)\delta_1=(H-zI)^{-1} \delta_1.
$$
By the definition of the resolvent of the operator $H$, we have

\begin{eqnarray}
\pp (n-1)+\pp (n+1) +(V(n)-z) \pp(n) & = & 0, \ \ n \ne 1,\\ 
\label{no1}
\pp(0)+\pp(2)+(V(1)-z) \pp (1) & = & 1
\end{eqnarray}
in the case of $\ell^2(\Z)$ and

\begin{eqnarray}
\pp(n-1)+\pp(n+1)+(V(n)-z) \pp (n) & = & 0, \ \ n>1, \\
\label{no2}
\pp (2)+(V(1)-z) \pp (1) & = & 1
\end{eqnarray}
in the case of $\ell^2(\N)$. Consider the vectors $\Phi (n)=(\pp (n+1), \pp (n))^T$. One can easily see that $\Phi (n)=A(n,z) \Phi(n - 1)$ for $n \ne 1$. Therefore,
 
\begin{equation}\label{f1}
\Phi(n)=T(n,1;z) \Phi (1), \ \ n>1
\end{equation}
in both cases. Moreover, in the case of $\ell^2(\Z)$, one has the identity

\begin{equation}\label{f2}
\Phi(n) = T(n,0;z) \Phi (0), \ \ n<0.
\end{equation}
Since ${\rm det} \, A(n,z)=1$, the same is true for $T(n,m;z)$, so that $\|T^{-1}(n,m;z)\|=\|T(n,m;z)\|$ for all $n,m,z$. Therefore, (\ref{f1}) and \eqref{f2} imply

\begin{equation}\label{f3}
\|\Phi(n)\|\ge \|T(n,1;z)\|^{-1}\|\Phi (1)\|, \ \ n>1
\end{equation}
in both cases and

\begin{equation}\label{f4}
\|\Phi(n)\|\ge \|T(n,0;z)\|^{-1} \|\Phi (0)\|, \ \ n<0,
\end{equation}
in the case of $\ell^2(\Z)$. It is clear that to get a lower bound for the resolvent, we need an upper bound for the norm of the transfer matrix for complex values of~$z$. Usually, one has
such bounds only for real values of $z$, so we should establish relations between these two cases. 

\begin{lemma}\label{first}
Let $E \in {\R}, N>0$. Define 
$$
K(N)=\sup_{|n|\le N, |m| \le N} \|T(n,m;E)\|
$$
in the case of $\ell^2(\Z)$ and 
$$
L(N)=\sup_{1 \le n,m \le N} \|T(n,m;E)\|
$$
in the case of $\ell^2(\N)$. Let $\delta \in \C$. The following bounds hold:
\begin{equation}\label{f5}
\|T(n,1;E+\delta)\| \le K(N) \exp (K(N) |n| |\delta|), \  1\le n \le N, \ \ 
\end{equation}
\begin{equation}\label{f6}
\|T(n,0;E+\delta)\| \le K(N) \exp (K(N) |n| |\delta|), \ -N \le n \le 0,
\end{equation}
in the case of $\ell^2(\Z)$ and
\begin{equation}\label{f7}
\|T(n,1;E+\delta)\| \le L(N) \exp (L(N) |n| |\delta|), \ 1 \le n \le N, \ \ 
\end{equation}
in the case of $\ell^2(\N)$. 
\end{lemma}

\begin{proof}
The proof is virtually identical to the proof of Theorem~2J in \cite{Simon}. For example, for any $n$ with $2 \le n \le N$, one can write the identity
$$
T(n,1;E+\delta)=T(n,1;E)+\delta \sum_{j=1}^{n-1} T(n,j+1;E)B T(j,1;E+\delta), 
$$
where
$$
B= \left( \begin{array}{cc} 1 & 0 \\ 0 & 0 \end{array} \right).
$$
By iteration, using the fact that $\|T(n,m;E)\| \le K(N)$ for all $1 \le n,m \le N$, one can show \eqref{f5}. The same proof yields \eqref{f7}. The  bound \eqref{f6} can be proved in a similar manner.
\end{proof}

\begin{proof}[Proof of Theorem~\ref{main}.]
We give the proof in the case of $\ell^2(\Z)$, the proof in the case of $\ell^2(\N)$ is analogous, with some aspects being even simpler than in the whole-line case.
 
The basic idea is the same as in \cite{Tcherem} in the case of a model with a sparse potential. First, using Parseval's identity \cite{RS4}, one can write for any $n$,

\begin{equation}\label{f10}
\frac{1}{T} \int_0^\infty | \exp (-itH) \delta_1 (n)|^2 \exp (- 2t/T) \, dt
=  \frac{\e}{2 \pi} \int_{\R} |R(E+i \e) \delta_1(n)|^2 \, dE,
\end{equation}
where $\e=1/T$. For  given $E,\e$, we shall write $\pp (n)=R(E+i \e)\delta_1(n)$. The bounds \eqref{f3} and \eqref{f4} allow us to bound $\|\Phi(n)\|^2=|\pp (n+1)|^2+|\pp (n)|^2$ from below if we find upper bounds for $\|T(n,1;E+i\e)\|$ or $\|T(n,0;E+i \e)\|$, provided that $\|\Phi(1)\|$ or $\|\Phi(0)\|$, respectively, is not too small. 

Let $T > 1$. Define $N \equiv N(T) = T^{1/(1+\alpha)}$. The assumption of the theorem gives
$$
K(N) = \sup_{|n|\le N, |m| \le N} \|T(n,m;E')\| \le C N^{\alpha},
$$
provided that $E' \in A(N)$. Let $E \in B(T)$. Then there exists $E' \in A(N(T))\equiv A(N)$
such that $|\beta|\equiv |E-E'| \le 1/T$. 

Using Lemma~\ref{first}, we obtain for all $n$ with $1 \le n \le N$,

\begin{equation}\label{f11}
\|T(n,1;E+i \e)\| \le CN^{\alpha} \exp (C(|\beta|+\e) N^{\alpha+1})=D N^{\alpha}, 
\end{equation}
where $D=C \exp (2C)$, since $\e=1/T, |\beta| \le 1/T$ and $n \le N = T^{1/(1+\alpha)}$. Similarly, for all $n$ with $-N \le n \le 0$,
 
\begin{equation}\label{f12}
\|T(n,0;E+i\e)\| \le D N^{\alpha}.
\end{equation}
It follows from \eqref{f3}--\eqref{f4} and \eqref{f11}--\eqref{f12} that

\begin{equation}\label{f14}
|\pp (n+1)|^2 +|\pp (n)|^2 \ge K N^{-2 \alpha} (|\pp (2)|^2+|\pp (1)|^2 ),
\end{equation}
for all $E \in B(T)$, $1 \le n \le N$, and
 
\begin{equation}\label{f15}
|\pp (n+1)|^2 +|\pp (n)|^2 \ge K N^{-2 \alpha} (|\pp (1)|^2+|\pp (0)|^2 ), 
\end{equation}
for all $E \in B(T)$, $-N \le n \le 0$. Here, $K > 0$ is some uniform constant. Now with
\eqref{f14}--\eqref{f15}, we can estimate, for any $E \in B(T)$,

\begin{eqnarray}
\nonumber
\sum_{|n| \ge N/2+1} (|\pp (n+1)|^2+|\pp (n)|^2) & \ge & \sum_{n = N/2+1}^N + \sum_{n = -N}^{-N/2-1}\\
\label{f16}
& \ge & B (N/2-2)N^{-2\alpha} (|\pp (0)|^2+2 |\pp (1)|^2+|\pp (2)|^2) \\
\nonumber
& \ge & D N^{1-2\alpha} (|\pp (0)|^2+|\pp (1)|^2+|\pp (2)|^2)
\end{eqnarray}
with uniform constant $D>0$. Here we have assumed that $T$ is large enough, so that $N > 8$, for example. We now use equation \eqref{no1} for the resolvent for $n=1$:
$$
\pp (2)+\pp (0)+(V(1) - E - i \e) \pp (1)=1.
$$
Since $E \in B(T), A(N) \subset [-K,K]$ and $\e=1/T \le 1$, we get
$$
|\pp (2)|+|\pp (1)|+|\pp (0)| \ge \frac{1}{|V(1)|+K+2}.
$$
Together with \eqref{f16} this gives

\begin{equation}\label{f17}
\sum_{|n| \ge N/2+1} (|\pp (n+1)|^2+|\pp (n)|^2) \ge \gamma N^{1-2 \alpha}, 
\end{equation}
where $\gamma >0$ is some uniform constant depending on $|V(1)|, K, D$. It is clear that the left-hand side of \eqref{f17} is bounded from above by
$$
2 \sum_{|n| \ge N/2} |\pp (n)|^2.
$$
Thus, we obtain the following lower bound: For every $E \in B(T)$,
$$
\sum_{|n| \ge N/2} |R(E + i \e)\delta_1 (n)|^2 \ge \frac{\gamma}{2} N^{1-2\alpha}.
$$
Therefore, 

\begin{equation}\label{f18}
\int_{\R} \sum_{|n| \ge N/2} |R(E+i \e)\delta_1(n)|^2 \, dE \ge 
\int_{B(T)} \frac{\gamma}{2} N^{1-2\alpha} \, dE = \frac{\gamma}{2} |B(T)| N^{1-2\alpha}.
\end{equation}
Summation over $\{ n: |n| \ge N/2\}$ in \eqref{f10} together with \eqref{f18} proves \eqref{f8}. Since
$$
|X|^p(t) \ge (N/2)^p \sum_{|n| \ge N/2} |\exp (-itH) \delta_1(n)|^2,  
$$
the bound \eqref{f9} follows immediately.

In the half-line case, the proof is simpler. The equation \eqref{no2} for the resolvent for 
$n=1$ gives the uniform lower bound for $\|\Phi(1)\|$, and using \eqref{f14}, we prove the two statements of the Theorem.
\end{proof}

\begin{proof}[Proof of Corollary~\ref{oneenergy}] 
One takes $A(N)=\{ E_0 \}$ for all $N$. Since $|B(T)|=2/T$, the result follows
directly from (\ref{f9}). 
\end{proof}

\begin{proof}[Proof of Corollary~\ref{coro2}]
One takes $A(N)=[E_0-N^{-1/\theta}, E_0+N^{-1/\theta}]$. One sees easily that \eqref{maincond} holds with $\alpha=0$. Therefore, $N(T)=T$ and $|B(T)| \ge |A(N(T))|=2T^{-1/\theta}$. The bound \eqref{f9} yields the result. 
\end{proof}

\begin{proof}[Proof of Corollary~\ref{coro3}]
It is easy to see that the operator $H$ satisfies the hypothesis of Theorem~\ref{main} with the same $K$, $\alpha$, $A(N)$, and some appropriately adjusted constant $C$. This yields \eqref{f8} and \eqref{f9}.
\end{proof}

%
%
%
%

\section{The Fibonacci Model}

In this section we apply Theorem~\ref{main} to the Fibonacci model. We refer the reader to \cite{d2} for background information on this model and related ones. While Theorem~\ref{main} already gives a non-trivial dynamical bound when the set $A(N)$ consists of a single point, and such an input is known for the Fibonacci model \cite{it}, we shall nevertheless identify the natural set $A(N)$ of energies for which one can prove the local power-law bounds required by Theorem~2. On the one hand, this improves the dynamical bound, since we obtain a larger lower bound for the measure of $B(T)$, and on the other hand it illustrates nicely how the sets, on which one can prove local power-law bounds (they will turn out to be spectra of periodic approximants) shrink down to a zero-measure set (the spectrum of the Fibonacci Hamiltonian) of energies for which the power-law bounds hold globally.

Let us first present some basic notions and results for the Fibonacci
potential given by \eqref{fibpot}. The transfer matrices $T(n,m;z)$ are
defined as in \eqref{transfermat}, and we shall sometimes write
$T_\lambda(n,m;z)$ to make their dependence on the parameter $\lambda$ explicit.

Define the sequence $(F_k)_{k \ge 0}$ of Fibonacci numbers by
$$
F_0 = 1, \; F_1 = 1, \; F_{k+1} = F_k + F_{k-1} \mbox{ for } k \ge 1.
$$
Let
$$
x_k (E,\lambda) = \tr \, M_k(E, \lambda),
$$
where
$$
M_k(E,\lambda) = T_\lambda(F_k,1;E).
$$
We will often leave the dependence of $x_k$ or $M_k$ on $E, \lambda$ implicit. The matrices $M_k$ obey the recursion

\begin{equation}\label{matrec}
M_k = M_{k-2} M_{k-1}
\end{equation}
which yields

\begin{equation}\label{tracemap}
x_{k+1} = x_k x_{k-1} - x_{k-2}
\end{equation}
for their traces. This in turn gives the invariant

\begin{equation}\label{invariant}
x_{k+1}^2 + x_k^2 + x_{k-1}^2 - x_{k+1} x_k x_{k-1} = 4 + \lambda^2 \; \mbox{ for every } k \in \N.
\end{equation} 
The identities \eqref{matrec}--\eqref{invariant} were proved by S\"ut\H{o} in \cite{s}.

For fixed $\lambda$, define 
$$
\sigma_k = \{ E \in \R : |x_k (E, \lambda)| \le 2\}.
$$
The set $\sigma_k$ is actually equal to the spectrum of the Schr\"odinger operator $H$ whose potential $V_k$ results from $V$ in \eqref{fibpot} by replacing $\alpha$ by $F_{k-1}/F_k$ (see \cite{s}). Hence, $V_k$ is periodic, $\sigma_k \subset \R$, and it consists of $F_k$ bands (closed intervals).

\begin{lemma}\label{locality}
For every $E \in \sigma_k$, we have

\begin{equation}\label{locbound}
|x_i(E,\lambda)| \le C_\lambda \; \mbox{ for } 0 \le i \le k.
\end{equation}
\end{lemma}

\begin{proof}
It was shown in \cite{s} that for every $m \in \N$,
$$
\sigma_m \cup \sigma_{m+1} \subseteq \sigma_{m-1} \cup \sigma_m.
$$
Thus, if $E \in \sigma_k$, then $|x_k(E,\lambda)| \le 2$ and for every $i < k$, we have that either $|x_i (E, \lambda)| \le 2$ or $\max \{ |x_{i-1}(E, \lambda)| , |x_{i+1}(E, \lambda)| \} \le 2$. In the former case, the claimed bound clearly holds. In the latter case, use the invariant \eqref{invariant} to establish the claimed upper bound for $|x_i(E, \lambda)|$.
\end{proof}

Lemma~\ref{locality} is the key tool in identifying a natural set of
energies for which the desired power-law bounds on transfer matrices up
to a certain distance from the origin hold. 

\begin{prop}\label{pl}
For every $\lambda$, there is a constant $C$ such that for every $k \in \N$, every $E \in \sigma_k$, and every $m$ with $-F_k + 1 \le m \le F_k$ {\rm (}$m \not= 0${\rm )}, we have

\begin{equation}\label{powerlaw}
\| T_\lambda(m,1;E) \| \le C |m|^\alpha,
\end{equation}
where

\begin{equation}\label{alphavalue}
\alpha = \frac{\log (C_\lambda (2 C_\lambda + 1)^2)}{\log \omega^{-1}}.
\end{equation}
\end{prop}

\begin{proof}
We first note that due to the symmetry $V(- n) = V(n - 1)$, $n \ge 2$ of the potential \cite{s}, we can restrict our attention to $1 \le m \le F_k$.

Our principal strategy is to modify the approach of \cite{it} as to treat a larger set of energies, while proving bounds only in finite regions. Essentially, their proof of a power-law upper bound for energies in the spectrum can be turned into the claimed result by noting that for each $m$, boundedness of traces is only needed for $j$'s with $F_j \le m$, and the previous lemma established this fact for energies in $\sigma_k$, where $k = \max\{ j : F_j \le m \}$.

For the reader's convenience, let us be more concrete. Fix $\lambda$, $k$, and $E \in \sigma_k$. By symmetry ($V(-n) = V(n-1)$, $n \ge 2$; see \cite{s}), we can restrict our attention to the case of positive $m$. The proof of \eqref{powerlaw} will be split into several steps. 

\textit{Step 1.} For $0 \le i \le k$, we have
 
\begin{equation}\label{normone}
\|M_i \| \le (C_\lambda)^i.
\end{equation}
This follows from
$$
M_i = M_{i-2} M_{i-1} = M_{i-2} ( x_{i-1} I - M_{i-1}^{-1} ) = x_{i-1} M_{i-2} - M_{n-3}^{-1}
$$
along with $\|M_{i-3}^{-1}\| = \| M_{i-3} \|$.

\textit{Step 2.} For $i,j \in \N$ with $i+j \le k$ and $i \ge 2$, we have

\begin{equation}
M_i M_{i+j} = P_j^{(1)} M_{i+j} + P_j^{(2)} M_{i+j-1} + P_j^{(3)} M_{i+j-2} + P_j^{(4)} I,
\end{equation}
where, for $l = 1,2,3,4$, $P_j^{(l)} = P_j^{(l)}(x_{i-1} , x_i , \ldots , x_{i+j})$ is a polynomial of degree at most $j$, and we have

\begin{equation}
\sum_{l = 1}^4 |P_j^{(l)}| ( |x_{i-1}| , \ldots , |x_{i+j}| ) \le (2C_\lambda + 1)^j,
\end{equation}
where the polynomial $|P_j^{(l)}|$ results from $P_j^{(l)}$ by replacing all the coefficients by their respective absolute values.

To see this, given \eqref{locbound}, one can literally redo the proof of Lemma~5 in \cite{it} since it only uses the trace bound for indices bounded by $k$.

\textit{Step 3.} For $1 \le m \le F_k$, we have

\begin{equation}\label{intbound}
\|T_\lambda(m,1;E)\| \le d^{m_N},
\end{equation}
where
$$
d = C_\lambda (2 C_\lambda + 1)^2
$$
and $m_N$ appears in the unique coding of $m$ in terms of the Fibonacci numbers,
$$
m = \sum_{l = 0}^N F_{m_l}, \; F_{m_0} < F_{m_1} < \cdots < F_{m_N}, \; m_l - m_{l-1} \ge 2.
$$
Clearly, $m_N = \max\{ i : F_i \le m \}$ and hence $m_N \le k$.

The estimate \eqref{intbound} can be proved in the exact same way as in \cite{it}, given Steps~1 and 2 above. 

\textit{Step 4.} Conclusion of the proof: The Fibonacci numbers $F_i$ behave asymptotically like $\omega^{-i} = [(1 + \sqrt{5})/2]^i$ and in particular, there is a constant $D_1$ such that
$$
F_i \ge D_1 \omega^{-i} \; \mbox{ for every } i \ge 1.
$$
This means that the number $m_N$ defined in Step~3 obeys
$$
m_N \le \frac{\log m}{\log \omega^{-1}} + D_2
$$
for some suitable constant $D_2$, which in turn implies
$$
\|T_\lambda(m,1;E)\| \le d^{\frac{\log m}{\log \omega^{-1}} + D_2},
$$
and hence \eqref{powerlaw} for a constant $C$ depending only on $\lambda$ and $\alpha$ as in \eqref{alphavalue}.
\end{proof}

We therefore have found a natural candidate that serves the purpose of the set $A(N)$ in Theorem~\ref{main}. Namely, given some $N$, choose $k$ such that $F_{k-1} < N \le F_k$ and set $A(N) = \sigma_k$.

Next, we prove a lower bound for the Lebesgue measure of $\sigma_k$. More precisely, we obtain lower bounds for each one of the $F_k$ intervals it is made of. The results presented here are essentially contained in Raymond \cite{r}. We give a somewhat more streamlined presentation in the spirit of \cite{kkl}.

From now on, we shall always assume

\begin{equation}\label{four}
\lambda > 4,
\end{equation}
since we will make critical use of the fact that in this case, it follows from the invariant \eqref{invariant} that three consecutive traces cannot be bounded in absolute value by $2$:

\begin{equation}\label{critical}
\forall \lambda > 4, \forall E,k : \max \{ |x_k(E,\lambda)|, |x_{k+1}(E,\lambda)|, |x_{k+2}(E,\lambda)| \} > 2.
\end{equation}

Following \cite{kkl}, we call a band $I_k \subset \sigma_k$ a type A band if $I_k \subset \sigma_{k-1}$ (and hence $I_k \cap (\sigma_{k+1} \cup \sigma_{k-2}) = \emptyset$). We call a band $I_k \subset \sigma_k$ a type B band if $I_k \subset \sigma_{k-2}$ (and therefore $I_k \cap \sigma_{k-1} = \emptyset$).

From \eqref{critical}, one gets the following (Lemma~5.3 of \cite{kkl}, essentially Lemma~6.1 of \cite{r}):

\begin{lemma}\label{order}
For every $\lambda > 4$ and every $k \in \N$, 

\begin{itemize}
\item[{\rm (a)}] Every type A band $I_k \subset \sigma_k$ contains exactly one type B band $I_{k+2} \subset \sigma_{k+2}$, and no other bands from $\sigma_{k+1}$, $\sigma_{k+2}$. 
\item[{\rm (b)}] Every type B band $I_k \subset \sigma_k$ contains exactly one type A band $I_{k+1} \subset \sigma_{k+1}$ and two type B bands from $\sigma_{k+2}$, positioned around $I_{k+1}$.
\end{itemize}
\end{lemma}

We will also need the following lemma (Lemma~5.4 of \cite{kkl}, essentially Proposition~A.2 of \cite{r}).

\begin{lemma}\label{partbound}
Let the functions $f_\pm (x,y,\lambda)$ be defined by
$$
f_\pm (x,y,\lambda) = \frac{1}{2} \left[ xy \pm \sqrt{ 4 \lambda^2 + ( 4 - x^2 ) ( 4 - y^2 ) } \right] .
$$
For $\lambda > 4$ and $-2 \le x,y \le 2$, we have
$$
\max \left\{ \left| \frac{\partial f_\pm}{\partial x} (x,y, \lambda) \right| , \left| \frac{\partial f_\pm}{\partial y} (x,y, \lambda) \right| \right\} \le 1.
$$
\end{lemma}

Equipped with the previous two lemmas, we are now in position to prove the following:

\begin{lemma}
For every $\lambda > 4$, the following holds:

\begin{itemize}
\item[{\rm (a)}] Given any {\rm (}type A{\rm )} band $I_{k+1} \subset \sigma_{k+1}$ lying in the band $I_k \subset \sigma_k$, we have for every $E \in I_{k+1}$,
$$
\left| \frac{x_{k+1}'(E)}{x_k'(E)} \right| \le \lambda + 11.
$$
\item[{\rm (b)}] Given any {\rm (}type B{\rm )} band $I_{k+2} \subset \sigma_{k+2}$ lying in the band $I_k \subset \sigma_k$, we have for every $E \in I_{k+2}$,
$$
\left| \frac{x_{k+2}'(E)}{x_k'(E)} \right| \le 2(\lambda + 11).
$$
\end{itemize}
\end{lemma}

\begin{proof}
(a) Differentiating \eqref{tracemap} and dividing by $x_k'$, we get

\begin{equation}\label{nice}
\frac{x_{k+1}'}{x_k'} = x_{k-1} + \frac{x_k x_{k-1}'}{x_k'} - \frac{x_{k-2}'}{x_k'}.
\end{equation}
Using the invariant \eqref{invariant} and Lemma~\ref{partbound}, we obtain for every $E \in I_{k+1}$,

\begin{eqnarray*}
|x_{k-1}| & = & |f_\pm (x_k,x_{k-2},\lambda)|\\
& = & \left| \frac{1}{2} \left( x_k x_{k-2} \pm \sqrt{4 \lambda^2 + ( 4 - x_k^2 )( 4 - x_{k-2}^2 ) } \right) \right| \\
& \le & \frac{1}{2} \left( 4 + \sqrt{4 \lambda^2 + 64} \right) \\
& \le & \lambda + 6.
\end{eqnarray*}
Lemma~\ref{partbound} implies

\begin{equation}\label{simple1}
|x_{k-1}'| = \left| \frac{\partial f_\pm}{\partial x} (x_k,x_{k-2},\lambda) x_k' + \frac{\partial f_\pm}{\partial y} (x_k,x_{k-2},\lambda) x_{k-2}' \right| \le |x_k'| + |x_{k-2}'|.
\end{equation} 
From Lemma~5.5 of \cite{kkl}, we infer, again for $E \in I_{k+1}$,

\begin{equation}\label{inverse}
\left| \frac{x_{k-2}'}{x_k'} \right| < 1.
\end{equation}
Using \eqref{nice}, \eqref{simple1}, and \eqref{inverse}, we get

\begin{eqnarray*}
\left| \frac{x_{k+1}'}{x_k'} \right| & \le & \lambda + 6 + |x_k| \frac{|x_{k-1}'|}{|x_k'|} + \frac{|x_{k-2}'|}{|x_k'|} \\
& \le & \lambda + 6 + 2 \left( 1 + \frac{|x_{k-2}'|}{|x_k'|} \right) + \frac{|x_{k-2}'|}{|x_k'|}\\
& \le & \lambda + 11.
\end{eqnarray*}
(b) We consider two cases. Assume first $I_{k+2} \cap \sigma_{k-1} = \emptyset$ (and so $I_{k+2} \subset \sigma_{k-2}$). Given
$$
x_{k+2}' = x_{k+1} x_k' + x_{k+1}' x_k - x_{k-1}'
$$
and
$$
x_{k+1}' = x_k x_{k-1}' + x_k' x_{k-1} - x_{k-2}',
$$
we find
$$
\frac{x_{k+2}'}{x_k'} = 2 x_{k+1} - x_{k-2} + (x_k^2 - 1) \frac{x_{k-1}'}{x_k'} - x_k \frac{x_{k-2}'}{x_k'}.
$$
Similarly to the previous argument,
$$
|x_{k+1}| \le \lambda + 6
$$
and
$$
|x_{k-1}'| \le |x_k'| + |x_{k-2}'|,
$$
which yields

\begin{eqnarray*}
\left| \frac{x_{k+2}'}{x_k'} \right| & \le & 2\lambda + 12 + 2 + 3 \left( 1 + \frac{|x_{k-2}'|}{|x_k'|} \right) + 2 \frac{|x_{k-2}'|}{|x_k'|} \\
& \le & 2 \lambda + 22.
\end{eqnarray*}
Let us now assume $I_{k+2} \subset \sigma_{k-1}$ (and so $I_{k+2} \cap \sigma_{k-2} = \emptyset$). We proceed analogously and obtain
$$
\frac{x_{k+2}'}{x_k'} = x_{k+1} + x_k \frac{x_{k+1}'}{x_k'} - \frac{x_{k-1}'}{x_k'},
$$
$$
|x_{k+1}| \le \lambda + 6,
$$
and
$$
|x_{k+1}'| \le |x_k'| + |x_{k-1}'|,
$$
which then yields

\begin{eqnarray*}
\left| \frac{x_{k+2}'}{x_k'} \right| & \le & \lambda + 6 + 2 \frac{|x_k'| + |x_{k-1}'|}{|x_k'|}  + \frac{|x_{k-1}'|}{|x_k'|} \\
& \le & \lambda + 11,
\end{eqnarray*}
concluding the proof.
\end{proof}

This yields, as a very rough estimate, that for $\lambda > 4$ and $E \in \sigma_k$, we have

\begin{equation}\label{derbound}
|x_k'(E,\lambda)| \le C (2\lambda + 22)^k,
\end{equation}
which can easily be turned into a lower bound on bandwidths, as the following proposition shows.

\begin{prop}\label{measure}
For every $\lambda > 4$, the set $\sigma_k$ consists of $F_k$ disjoint closed intervals, each of which has Lebesgue measure bounded from below by

\begin{equation}\label{bandbound}
|I_k| \ge \frac{4}{C (2\lambda + 22)^k}.
\end{equation}
In particular, we obtain

\begin{equation}\label{measbound}
|\sigma_k| \ge \frac{4}{C} F_k^{- \gamma},
\end{equation}
where $\gamma$ is as in \eqref{gammalambda}.
\end{prop}
 
\begin{proof}
Since $\sigma_k$ is the spectrum of a periodic Schr\"odinger operator with period $F_k$, the first statement is immediate. That the gaps are open was shown in \cite{r}. The estimate \eqref{bandbound} for the measure of one of these $F_k$ bands can be seen as follows (cf.~\cite{r}): On such a band $I_k$, $x_k(E,\lambda)$ runs monotonically from $\pm 2$ to $\mp 2$. Hence, by \eqref{derbound},
$$
4 = \int_{I_k} | x_k'(t,\lambda) | \, dt \le |I_k| \cdot C (2\lambda + 22)^k.
$$ 
The estimate \eqref{measbound} then follows from this and the exponential behavior of the sequence $(F_k)_{k \in \N}$.
\end{proof}

We have established the input to Theorem~\ref{main}. Note that in any event, we have $|B(T)| \ge 2/T$. In the case where $\lambda > 4$, Proposition~\ref{measure} improves this factor in \eqref{f8} and \eqref{f9}. We therefore proceed with the

\begin{proof}[Proof of Theorem~\ref{fibtheo}.]
(a) This follows from Proposition~\ref{pl} and Corollary~\ref{oneenergy}.\\
(b) Given $T$, we let as above $N(T) = T^{1/(1+\alpha)}$ and we choose $k$ such that $F_{k-1} < N(T) \le F_k$. We let $A(N(T)) = \sigma_k$. Thus, from Proposition~\ref{measure} we get
$$
|B(T)| \ge \frac{4}{C} F_k^{-\gamma} \ge \frac{4}{C} (2 F_{k-1})^{-\gamma} \ge \frac{4}{C} 2^{-\gamma} N(T)^{- \gamma} = \frac{4}{C} 2^{-\gamma} T^{\frac{- \gamma}{1 + \alpha}}.
$$
Now \eqref{f9} yields 
$$
\langle |X|_\psi^p \rangle (T)  \ge \tilde{C} 2^{- \gamma} T^{\frac{p - \gamma - 3 \alpha}{ 1 + \alpha}}
$$
for $T$ large enough and hence $\beta^- (p) \ge (p - \gamma - 3 \alpha)/( 1 + \alpha)$.
\end{proof}

%
%
%
%

\section{The Period Doubling Model}

In this section we investigate the period doubling model and, in particular, prove Theorem~\ref{pdtheo}. This will be done by first proving linear bounds on transfer matrix norms for a finite/countable set of energies and then alluding to Corollary~\ref{oneenergy}.

On the alphabet $A=\{0,1\}$, consider the period doubling substitution $S(0)=01$, $S(1)=00$. Iterating on $0$, we obtain a one-sided sequence $u = 01000101 \ldots$ which is invariant under the substitution process. Define the associated subshift $\Omega_{{\rm pd}}$ to be the set of all two-sided sequences over $A$ which have all their finite subwords occurring in $u$. For $\lambda \in \R$ and $\omega \in \Omega_{{\rm pd}}$, we define the potential $V_{\lambda,\omega}$ by $V_{\lambda,\omega} (n) = \lambda \omega_n$.

If $w = w_1 \ldots w_l$ with $w_i \in \{0,1\}$, we define
$$
T_\lambda(w;E) = A_\lambda(w_l;E) \times \cdots \times A_\lambda(w_1;E),
$$
where for $a \in \{0,1\}$,
$$
A_\lambda(a;E) = \left( \begin{array}{cr} E - \lambda a & -1 \\ 1 & 0 \end{array} \right).
$$
We let
$$
T^{(0)}_k = T^{(0)}_k (E,\lambda) = T_\lambda(S^k(0);E), \; T^{(1)}_k = T^{(1)}_k (E,\lambda) = T_\lambda(S^k(1);E).
$$
In the following, we will leave the dependence of $T^{(0)}_k$, $T^{(1)}_k$ on $E$ and $\lambda$ implicit. We also define 
$$
x_k = {\rm tr} \, T^{(0)}_k , \; y_k = {\rm tr} \, T^{(1)}_k.
$$
It follows from the substitution rule (and is easy to check) that
$$
T^{(0)}_{k+1} = T^{(1)}_k T^{(0)}_k , \; T^{(1)}_{k+1} = T^{(0)}_k T^{(0)}_k
$$
and 

\begin{equation}\label{pdtm}
x_{k+1} = x_k y_k - 2 , \; y_{k+1} = x_k^2 - 2.
\end{equation}
The relation \eqref{pdtm} is called the period doubling trace map.

By virtue of Corollary~\ref{oneenergy}, Theorem~\ref{pdtheo} follows once we find an energy $E_0$ such that \eqref{oneecond} holds with $\alpha = 1$. We establish this first and then discuss later what the natural set of energies is for which one can establish a bound like \eqref{oneecond}.

\begin{proof}[Proof of Theorem~\ref{pdtheo}.]
Clearly, we have
$$
T^{(0)}_0 = \left( \begin{array}{cr} E & -1 \\ 1 & 0 \end{array} \right)
$$
and
$$
T^{(1)}_0 = \left( \begin{array}{cr} E - \lambda & -1 \\ 1 & 0 \end{array} \right).
$$
Notice that for $E_0 = 0$, we have
$$
T^{(0)}_0 = \left( \begin{array}{cr} 0 & -1 \\ 1 & 0 \end{array} \right) , \; T^{(1)}_0 = \left( \begin{array}{rr} -\lambda & -1 \\ 1 & 0 \end{array} \right).
$$
Thus, for this choice of the energy, we get
$$
T^{(0)}_1 = T^{(1)}_0 T^{(0)}_0 = \left( \begin{array}{rr} -\lambda & -1 \\ 1 & 0 \end{array} \right) \cdot \left( \begin{array}{rr} 0 & -1 \\ 1 & 0 \end{array} \right) = \left( \begin{array}{rr} -1 & \lambda \\ 0 & -1 \end{array} \right)
$$
and
$$
T^{(1)}_1 = T^{(0)}_0 T^{(0)}_0 = \left( \begin{array}{rr} 0 & -1 \\ 1 & 0 \end{array} \right) \cdot \left( \begin{array}{rr} 0 & -1 \\ 1 & 0 \end{array} \right) = \left( \begin{array}{rr} -1 & 0 \\ 0 & -1 \end{array} \right)
$$
In particular, $T^{(0)}_1$ and $T^{(1)}_1$ commute. Notice that
$$
\left( T^{(0)}_1 \right)^n = (-1)^n \left( \begin{array}{rc} 1 & - n \lambda \\ 0 & 1 \end{array} \right).
$$
Every subword $w$ of $u$ can be partitioned into a product of blocks of the form $S(a)$ or $S(b)$, up to a possible prefix/suffix of length one. We therefore get, for every $\omega \in \Omega_{{\rm pd}}$,
$$
\|T(n,m; E_0)\| \le C_\lambda^2 \left( \sqrt{2} + \frac{\lambda}{2} |n-m| \right),
$$
where 
$$
C_\lambda = \left\| \left( \begin{array}{rr} -\lambda & -1 \\ 1 & 0 \end{array} \right) \right\| \le \sqrt{2} + \lambda.
$$
We can now apply Corollary~\ref{oneenergy}.
\end{proof}

We now study the set of exceptional energies (i.e., where we get commuting transfer matrices at some level) a little further. We prove the following:

\begin{lemma}
Fix some coupling constant $\lambda$ and some $k \in \N_0$. If $E$ is such that $x_n = 0$, then there is a constant $a_{\lambda,k}$ such that

\begin{equation}\label{t0n}
T^{(0)}_{k+1} \mbox{ is conjugate to } \left( \begin{array}{rc} -1 & a_{\lambda,k} \\ 0 & -1 \end{array} \right)
\end{equation}
and 

\begin{equation}\label{t1n}
T^{(1)}_{k+1} = \left( \begin{array}{rr} -1 & 0 \\ 0 & -1 \end{array} \right).
\end{equation}
\end{lemma}

\begin{proof}
If $x_k = 0$, then \eqref{pdtm} gives $x_{k+1} = -2$. Hence, \eqref{t0n} is immediate. Moreover, \eqref{t1n} follows from the Cayley-Hamilton theorem since we have
$$
T^{(1)}_{k+1} = T^{(0)}_k T^{(0)}_k = x_k \cdot T^{(0)}_k - {\rm Id} = \left( \begin{array}{rr} -1 & 0 \\ 0 & -1 \end{array} \right).
$$
\end{proof}

This allows us to prove:

\begin{prop}
Fix some coupling constant $\lambda$ and some $k \in \N_0$. For every root $E_0$ of $x_k$, we have
$$
\|T(n,m; E_0)\| \le C_{\lambda,k}^2 \left( \sqrt{2} + \frac{a_{\lambda,k}}{2^k} |n-m| \right),
$$
where $C_{\lambda,k}$ is some suitable constant.
\end{prop}

\begin{proof}
The argument is virtually the same as the one used in the proof of Theorem~\ref{pdtheo}.
\end{proof}

Notice that $x_k$ is, as a function of $E$, a polynomial of degree $2^k$ which has exactly $2^k$ roots. In particular, there is a countable set of energies where linear bounds on the growth of transfer matrix norms can be proved.

%
%
%
%

\section{The Thue-Morse Model}

In this section we investigate the Thue-Morse model and, in particular, prove Theorem~\ref{tmtheo}. This will be done by exhibiting a set of energies for which the transfer matrix norms are bounded and then alluding to Corollary~\ref{oneenergy}.

On the alphabet $A=\{0,1\}$, consider the Thue-Morse substitution $S(0)=01$, $S(1)=10$. Iterating on $0$, we obtain a one-sided sequence $u = 01101001 \ldots$ which is invariant under the substitution process. Define the associated subshift $\Omega_{{\rm tm}}$ to be the set of all two-sided sequences over $A$ which have all their finite subwords occurring in $u$. For $\lambda \in \R$ and $\omega \in \Omega_{{\rm tm}}$, we define as in the period doubling case the potential $V_{\lambda,\omega}$ by $V_{\lambda,\omega} (n) = \lambda \omega_n$.

We can now define transfer matrices in the same way as above, that is,
$$
T^{(0)}_k = T^{(0)}_k (E,\lambda) = T_\lambda(S^k(0);E), \; T^{(1)}_k = T^{(1)}_k (E,\lambda) = T_\lambda(S^k(1);E).
$$
Again, we will leave the dependence of $T^{(0)}_k$, $T^{(1)}_k$ on $E$ and $\lambda$ implicit. We define 
$$
x_k = {\rm tr} \, T^{(0)}_k , \; y_k = {\rm tr} \, T^{(1)}_k.
$$
It is clear that $x_k = y_k$ for $k \ge 1$ and it follows from the substitution rule that
$$
T^{(0)}_{k+1} = T^{(1)}_k T^{(0)}_k , \; T^{(1)}_{k+1} = T^{(0)}_k T^{(1)}_k
$$
and 

\begin{equation}\label{tmtm}
x_{k+1} = x_{k-1}^2 (x_k - 2) + 2 \; \mbox{ for } k \ge 2.
\end{equation}
The relation \eqref{tmtm} is called the Thue-Morse trace map.

By virtue of Corollary~\ref{oneenergy}, Theorem~\ref{tmtheo} follows once we find an energy $E_0$ such that \eqref{oneecond} holds with $\alpha = 0$. Let $\mathcal{E}_k = \{ E : x_k = 2 \}$.

\begin{prop}\label{tmkey}
If $k \ge 3$ and $E \in \mathcal{E}_k \backslash \mathcal{E}_2$, then

\begin{equation}\label{ident}
T^{(0)}_k = T^{(1)}_k = \left( \begin{array}{rr} 1 & 0 \\ 0 & 1 \end{array} \right).
\end{equation}
\end{prop}

\begin{proof}
This can be extracted from \cite{ap}. For the reader's convenience, we give a short proof of this fact. Notice that it follows from \eqref{tmtm} that for $k \ge 3$, we have $x_k = 2$ if and only if $x_{k-1} = 2$ or $x_{k-2} = 0$. Iterating this, we get that $x_k = 2$ holds if and only if $x_2 = 2$ or $x_j = 0$ for some $1 \le j \le k-2$. Thus, if $k \ge 3$ and $E \in \mathcal{E}_k \backslash \mathcal{E}_2$, then $x_j = 0$ for some $1 \le j \le k-2$. Using this and the Cayley-Hamilton theorem, we obtain

\begin{eqnarray*}
T^{(0)}_{j+2} & = & T^{(0)}_j T^{(1)}_j T^{(1)}_j T^{(0)}_j \\
& = & T^{(0)}_j \left( x_j T^{(1)}_j - I \right) T^{(0)}_j \\
& = & - T^{(0)}_j T^{(0)}_j \\
& = & - \left( x_j T^{(0)}_j - I \right) \\
& = & I
\end{eqnarray*}
and, similarly, $T^{(1)}_{j+2} = I$. This yields \eqref{ident}.
\end{proof}

\begin{proof}[Proof of Theorem~\ref{tmtheo}.]
Given Proposition~\ref{tmkey}, the claim follows as in the proof of Theorem~\ref{pdtheo} from Corollary~\ref{oneenergy}.
\end{proof}

For example, the reader may verify that if $E$ is chosen such that $E (E - \lambda) = 2$, then \eqref{ident} holds for $k = 3$ by a straightforward calculation. This observation already suffices for an application of Corollary~\ref{oneenergy}.

We conclude with a remark about the special energies exhibited by Proposition~\ref{tmkey}. It follows from \eqref{tmtm} that for $k \ge 2$, $\mathcal{E}_k \subset \mathcal{E}_{k+1}$. Axel and Peyri\`{e}re show that the the union of the sets $\mathcal{E}_k$ is dense in the spectrum of the Thue-Morse Hamiltonian \cite{ap}.

\end{document}